\documentclass[preprint,12pt]{aastex}
\usepackage{graphicx}
\usepackage{float}
\usepackage[space]{grffile}
\usepackage{latexsym}
\usepackage{amsfonts,amsmath,amssymb}
\usepackage{url}
\usepackage[utf8]{inputenc}
\usepackage{fancyref}
\usepackage{amsmath}
\usepackage[dvips]{color}
\newcommand\numberthis{\addtocounter{equation}{1}\tag{\theequation}}
\def\mspsr{millisecond pulsar}

 \def\ns{neutron star}

\def\hz{habitable zone}

\def\pl{planet}
\def\ps{planetary system}

\def\ac{advanced civilization}

\def\cc{communicating civilization}

\def\ffp{free-floating planet}
\def\Ffp{Free-floating planet}

\def\sfr{star-forming region}

\def\gc{globular cluster}
\def\Gc{Globular cluster}
\def\GC{Globular Cluster}

\def\mty{metalicity}
\def\mts{metalicities}

\def\gco{globular cluster opportunity} 
\shorttitle{Advanced Civilizations in Globular Clusters}
\shortauthors{Di Stefano \& Ray}

\begin{document}

\title{
Globular Clusters as Cradles of Life and Advanced Civilizations}

\author{R. Di\thinspace Stefano\\
Harvard-Smithsonian Center for Astrophysics\\ 
rdistefano@cfa.harvard.edu\\
\and
A. Ray\\
Tata Institute of Fundamental Research\\
akr@tifr.res.in}  

\begin{abstract}
Globular clusters are ancient stellar populations,
bound together in compact, dense ellipsoids. 
Because their stars are old, they are of low mass, and can burn 
steadily over many billions or trillions of years. 
There is little gas and dust,
hence no star formation or core-collapse supernovae. 
Although only a single
globular-cluster \pl \ has been discovered, evidence suggests that  
\gc s are rich in planets. If so, and if advanced civilizations
can develop in a \gc , then the distances between these civilizations
and other stars would be far smaller 
than typical distances between stars in the Galactic disk. 
The relative proximity would facilitate
interstellar communication and travel. 
We will refer to the potent combination of the 
long-term stability of \gc s and their
high stellar densities, as the {\sl \gc\ opportunity.}
However, the very proximity that promotes interstellar 
travel also  
brings danger, since stellar interactions can destroy 
planetary systems. 
Fortunately, we find that
large regions of many \gc s can be thought of as ``sweet spots''
where habitable-zone \pl ary orbits can be stable for long times.   
We use a Plummer model to compute the
ambient densities and fluxes in the regions within which habitable-zone
\pl s can survive, to verify that the \gco \ is real. 
Overall, \gc s in our own and other galaxies
are among the best targets for
searches for extraterrestrial intelligence (SETI). We
use the Drake equation to  compare \gc s
to the Galactic disk, in terms of the
likelihood of housing advanced communicating civilizations.
We also consider free-floating planets, since wide-orbit
\pl s can be ejected and travel freely through the cluster.
A civilizations spawned in a \gc \ may  have opportunities to
establish 
self-sustaining outposts, thereby reducing the 
probability that a single catastrophic event will destroy the civilization
or its descendants. Although individual civilizations
within a cluster may follow different evolutionary paths,
 or even be destroyed, the cluster may always host some advanced civilization,
once a small number of them have managed to jump across interstellar
space. Civilizations residing in \gc s could therefore, 
 in a sense, be immortal. 
\end{abstract}
\section{Introduction}

\Gc s are among the most ancient bound stellar systems in the Universe.
They contain $\sim 10^5$ to more than $10^6$ stars
in dense spheroids. Typical ages of 
the $\sim 150$ \gc s in the Milky Way are larger than $10$~Gyr,
extending to $\sim 13$~Gyr. [See, e.g., \citet{2015ApJ...812...25M},
\citet{2015AJ....150..155K} and references therein.]    
Here we consider the possibility that \gc s host \pl s, and that life
and advanced civilizations 
can develop and evolve there. Such civilizations would
be immersed in stellar environments so dense that 
distances between stars could be as small as hundreds or 
thousands of AU: thousands to hundreds of
times smaller than typical interstellar distances in the 
Milky Way's disk, which is home to the Sun. 
Interstellar communication between neighboring stars
could take as little as 
weeks to months, and only decades from the center of the cluster to
its edges. At a time when astronomical tools and techniques are  
as developed as those we now have on Earth,
most of the planets
within the cluster would have been discovered, and the large numbers
of photons incident from cluster stars would allow many detailed
studies of  exoplanetary
atmospheres.   
\Gc\ civilizations which reach a level of technical development
comparable to our own at present, will therefore 
know a good deal about the 
nearest $\sim 10^5$ stars and an even larger number of \pl s. They will be
able to send exploratory probes to nearby stars  
and receive data that takes only days, weeks, or months
to reach them. 
They may be able to travel to nearby \ps s  that are hospitable and establish
self-sustaining colonies over time scales far shorter than seems possible for 
\ac s which,  like our own, inhabit the relatively diffuse Galactic
disk. 
Independent outposts would
increase the chances of surviving
threats, 
ranging from astronomical and geological events to
civil strife.

We will refer to the potent combination of the 
long-term stability of \gc s and their
high stellar densities, as the {\sl \gc\ opportunity.}
In order for the \gco\ to be meaningful, planets must exist in \gc s.
We show in \S 2 that there are good reasons to expect that \gc s do
harbor populations of \pl s.
Planets in globular clusters, however, face threats of a type rarely
encountered in the Galactic disk.
Because of the high ambient stellar densities,
interactions with other stars 
are common and they are more likely to be
ejected
from their \ps s
or else
captured into the planetary systems of other stars.
Fortunately, not all orbital separations are equally
dangerous. 

We show in \S 3 that 
\gc s can have large regions within which the following
conditions are satisfied.

\noindent{\bf (1)}~The orbits 
of habitable-zone \pl s
are stable with respect to interactions with passing stars.
These regions correspond the the {\sl \gc\ \hz s} ({GC-HZs}),
which may be thought of as extensions of the Galactic Habitable Zone 
(GHZ), the regions within a galaxy where life may exist   
[see, e.g.
\citet{2011AsBio..11..855G}]. 

\noindent{\bf (2)}~Nearest-neighbor distances are small. 
The reason for this criterion is to allow for short
interstellar communication and travel times. Here we will
focus on situations in which nearest-neighbor distances are smaller than 
$10^4$~AU (\S 3.5).  

\noindent We refer to regions satisfying both of these conditions
as 
``sweet spots''. 
We conclude \S 3 by connecting these considerations
to the existence and likely locations of \ffp s,
which may dominate the number density in \gc s. 
 
In \S 4 we 
turn to the issue of what the \gco\ means for the long-term survivability
of any \ac s that develop within \gc s. To do this we use the Drake equation to
compare conditions within \gc s with those in the Galactic disk.  
Section \S 5 is devoted to an overview, 
a general discussion
of the implications of the \gco \ 
for future searches for planets and for
advanced extraterrestrial civilizations, 
and to the identification of \gc s that
may be ideal places to search.

\section{Planets in \GC s}

Galactic \gc s are old and their stars tend to  
have low \mts \ \citep{2010yCat..74011965H}.
Because \pl\ formation requires metals,
it could have been the case that \pl s did not form in \gc s.
Indeed, a null result was derived by a search for \pl s  in the \gc\ 47~Tuc.  
\citet{2000ApJ...545L..47G} studied $\sim 34,000$
main-sequence stars in 47~Tuc
to discover and measure the frequency
of ``hot Jupiters", gas giant planets in close orbits with their stars.
If the frequency of hot Jupiters in the 
observed portion of 47~Tuc, near its center, 
is the same as in the Solar neighborhood (about $1\%$), then  
this set of observations
should have detected $\sim 17$
planets with radii $\sim 1.3\, M_J$ and
typical orbital periods of 3.5 days.
No planet was detected.
This suggests that Jupiter-like \pl s in close orbits  
are
ten times less common in the center of
47 Tuc. This however does not place 
limits on Jovian planets in wider orbits
or on planets with radii substantially
smaller than Jupiter's.

Two different effects could have been responsible for this dearth of
hot Jupiters. First, because the central field is dense,
stellar interactions may have eliminated hot-Jupiter systems.
Studies of the outer, less dense, regions of 47~Tuc, also
failed to discover hot Jupiters \citep{2008ASPC..398..133W}, 
suggesting that dynamical interactions
were not the culprits. This was validated by dynamical simulations
that found that, had there been hot Jupiters in the
fields observed by Gilliland et al., they would have 
survived \citep{2006ApJ...640.1086F}.
The second effect is metalicity.   
An analysis of a sample of about $700$ exoplanets 
\citep{2012A&A...543A..45M}  found 
that the frequency of hot Jupiters declines with declining \mty .

Of particular interest to this investigation are planets
in the \hz s of low-mass stars,  
because the majority of \gc\ stars are M dwarfs. 
Through a detailed study of the {\sl Kepler} data,
taking into account detectability and selection effects,
\citet{2015ApJ...807...45D} estimate
that as many as one in four M-dwarf habitable
zones hosts an
Earth-sized planet, i.e., a planet of radius $1\, R_\Earth -1.5\, R_\Earth$.
In addition, approximately one in 5 M-dwarf habitable
zones hosts a super-Earth ($1.5\, R_\Earth -2\, R_\Earth$).
Because the frequency of low-mass \pl s does not follow the 
metallicity correlation found for hot Jupiters
\citep{2012Natur.486..375B,  2014Natur.509..593B, 2015ApJ...808..187B},  
the same statistics may apply to \gc s.

Thus, studies of metalicity effects in the
 field indicate that \pl s can form in \gc s,
and in the \hz s of their host stars. 
It is worth noting  
 that 
 the range of host-star \mts \ has significant overlap 
with the range of \mts \ measured for \gc s. 
Of 1709 \pl s listed in exoplanets.eu as of 6 December 2015, 927
have $z < 0$, and 278 have $z < -0.25$.
Of 134 \gc s with measured \mty\ from the Harris catalog,
28 are more metal-rich than 47~Tuc, which has a \mty\ of -0.76.
More than half of these higher-$z$ systems have \mty \ larger than -0.5, and
one is positive. 
Furthermore, some \gc s exhibit multiple stellar populations, each apparently
corresponding to a
slightly different formation time, with stars formed at later times
having higher metalicities \citep{2015ApJ...815L...4G}.

Because of the high ambient stellar densities,
\gc\ \pl s are more likely to be
ejected
from their \ps s
or else
captured into the planetary systems of other stars.
Nevertheless, Meibom et al. (2013) reported the discovery of 
planets smaller than Neptune in the old ($\sim 1$~Gyr) open cluster NGC6811. 
This example, and other recent planet discoveries in open 
clusters \citep{2012ApJ...756L..33Q, 2014A&A...561L...9B}, 
 show that \pl s can form and planetary orbits can survive
in dense environments, in spite of truncated protoplanetary
disks found in some clustered
environments 
\citep{2012A&A...546L...1D} 
and the relative fragility of some \ps s  in these environments
\citep{2015MNRAS.451..144P}.   
In addition, not all orbital separations are equally
dangerous.
We will show in \S 3 that
\gc s contain large ``sweet spots'', where 
planets in the habitable zones of low-mass stars can
survive for many Hubble times.

Because \gc s present crowded fields of dim stars,
traditional methods to search for planets don't yet do as 
well within \gc s as in the field.  
\Gc s do have one advantage, however, which is that the high interaction rates
produce low-mass X-ray binaries (LMXBs) that then morph 
into \mspsr s. 

The precise timing
of the pulses allows planets to be discovered through studies of the
residuals.
The \gc\ M4 contains
PSR B1620-26, with a spin period of 11 ms (and mass $\sim 1.35 \; M_{\odot}$).  
The pulsar is  part of a triple system with a
planetary mass object of $1-2 \; M_J$ orbiting a
neutron star-white 
dwarf binary system \citep{1993ApJ...412L..33T, 2003ApJ...597L..45R,
2003Sci...301..193S}. 
The white dwarf companion of the inner binary containing the neutron star
has a mass of $0.34 \pm 0.04 \; M_{\odot}$ in a low eccentricity
$e \sim 0.025$ orbit. It is a young white dwarf,
of age $\sim 0.5\, {\rm Gyr}$.
From pulsar timing limits the planet has a 45 yr
orbit with eccentricity ($e \sim 0.16$) with a semi-major
axis of $\sim 25 \rm AU$. The probability that a \ns\ will interact with
a particular \gc\ star is very low and is not significantly increased by
the presence of a planet. 
The discovery of this one planet must therefore signal the
existence of a large population of planets in M4.
This planet therefore demonstrates that \pl s may be common, even in a
\gc\ with $z< -1$, and even in a \gc\ with an interaction probability
high enough to produce a millisecond pulsar. 

\section{The Habitable Zone and the ``Sweet Spot''} 

\subsection{Survival in the Habitable Zone}

A star's habitable zone is defined to be the region around it within 
which planets like Earth can sustain water in liquid form 
on their surfaces. 
[For low-mass stars 
see, e.g., \citet{2007AsBio...7...30T}, \citet{2007AsBio...7...85S}.] 
Although the planet's atmosphere also plays a role in setting the surface 
temperature, a convenient and reasonable approximation to the
radius, $a$, of a habitable-zone orbit is:
$a = \sqrt{L/L_\odot} \, {\rm AU}$, where $L$ is the luminosity of the
star. For each star, there is a range of orbits in the \hz, and we will use
this expression as a guideline.    

Because the luminosities of stars decline steeply with
decreasing mass, 
the habitable zones of low-mass stars can have radii
considerably smaller than an AU.
Small orbits are more stable with respect to stellar interactions,
so that habitable-zone planets can have orbits that are stable over
long time intervals. 
Simulations of interactions between planetary systems and passing 
stars have been done by several groups. Their work provides estimates 
of the value of $\tau.$  We have found it convenient to use  
an expression taken
from \citet{2009ApJ...697..458S}:
$\tau = \frac{3\, v}{40\, \pi \, G\, a\, \rho}$, where 
$\rho$ is the mass density and $v$ the ambient speed. 
This expression is well suited for evaluation in a
simple cluster model
and is 
in approximate 
agreement with simulations  
by \citet{2006ApJ...640.1086F}.

More work is needed to
incorporate the full range of effects that play roles in
determining lifetimes. These effects include interactions
between \ps s with multiple \pl s and with stellar
systems that may be binaries or
higher order multiples. Below we give a simple
derivation showing that the uncertainty can be
incorporated into a factor that is likely to change by
less than an order of magnitude.

Consider a planetary system in which the stellar mass is $m_\ast$ and
the planet is in a circular orbit with radius $a$.
The rate $R$ at which stars pass within a distance $s$ of this \pl \
is dominated by gravitational focusing: 
$R= \pi\, {n}\, v\, s\, (2\, m_t\,G/v^2)$. Here $v$ is the
local average relative speed and ${n}$ is the local number density.
The combined mass of the two stars passing each other is $m_t$.

In order for a passage to significantly   
alter the orbit of the planet, leading for example to an ejection,
exchange, or merger, the impact parameter $s$ must be
comparable to $a$: $s = f\, a,$ where typically $f\approx 5 - 10.$ 
 The time between such close approaches gives an estimate of
the orbital lifetime, which is a function of $a$: $\tau(a)=1/R(a).$  
\begin{align*}
\tau
& = \frac{v}{2\, \pi \, (10\, a\ \times f/10)\,  m_t\,  G \, n} \\  
&  \\
& =  3 \times 10^8~{\rm yr}\, 
     \Big(\frac{v}{20\, {\rm km~s}^{-1}}\Big)\,   
     \Big(\frac{10^6\, {\rm pc}^{-3}}{n}\Big)\,   
     \Big(\frac{0.1\, {\rm AU}}{a}\Big)\,   
     \Big(\frac{0.5\, M_\odot}{m_t}\Big)\,   
     \Big(\frac{10}{f}\Big)\\   
& \\
& = 4 \times 10^9~{\rm yr}\,  
     \Big(\frac{v}{20\, {\rm km~s}^{-1}}\Big)\,   
     \Big(\frac{10^6\, {\rm pc}^{-3}}{n}\Big)\,   
     \Big(\frac{{0.2\, M_\odot}}{m_\ast}\Big)^{1.15}\,   
     \Big(\frac{0.5\, M_\odot}{m_t}\Big)\,   
     \Big(\frac{10}{f}\Big) \numberthis \\   
\end{align*}
The second line expresses $\tau$ in terms of $a$. The wider the
orbit, the shorter its lifetime. In this expression and the one just below,
the number density is scaled to a high value. The density in most portions 
of globular clusters is not this high, leading to longer planetary-orbit
lifetimes. For each cluster, the density decreases as the distance from the
center increases, making the lifetimes longer in the outer portions of the 
cluster.

Our focus is on planets in the \hz s of their stars. 
On the third line of Eq.~1 we utilize the mass-luminosity relationship for
the lowest mass stars [$L_\ast \approx 0.23\, (m/M_\odot)^{2.3}$].
The lifetime of 
planetary systems is longest for planets orbiting the least massive
stars. Interestingy, these stars also have very long lifetimes.
Hence there is a kind of serendipity: the stars which can provide the most
stable environments for life and evolution, can also 
harbor \pl s in \hz s that are 
relatively safe.

\subsection{Searching for a Globular Cluster's ``Sweet Spot''}
 
What we are seeking is a kind of ``sweet spot'' in the cluster, where 
habitable-zone orbits are stable, but the density of stars is still
large enough that interstellar travel can take less time. 
These 
two requirements are at odds with each other, 
since $\tau \propto 1/n,$
with large $\tau$ preferred, 
and $D\propto 1/n^\frac{1}{3}$, with small $D$ preferred. 

We expect the sweet spot to be a spherical shell that
starts at some distance $R_{low}^{sweet}$ from the cluster center
and ends at a larger radius $R_{high}^{sweet}$.
We will see that
clusters which have low central densities and which
are not highly concentrated will have 
sweet spots that start near the cluster center (small 
$R_{low}^{sweet}$), because survival
times may be long even there. But in such clusters, the fall
off of density with distance from the center will mean that
interstellar distances become large for stars far from the
center, so that the value 
of $R_{high}^{sweet}$ could be significantly smaller than the cluster's 
radius.  For clusters that are more concentrated, the sweet spot 
starts at larger distances from the center and may continue almost
to 
the cluster's outer edge. 
Thus, increasing concentration tends to move the sweet spot out. 

The concept of a sweet spot, is similar to the concept of a stellar 
habitable zone, or to a GHZ 
These concepts are useful in 
identifying regions that are most likely to harbor life. 
But their boundaries are not sharp, and there is
some arbitrariness in how we choose to define them. In the graphs illustrating
the results derived below, we have selected the sweet spots to 
begin at that distance from the cluster center where the survival
time of a habitable-zone planet is equal to the age of 
present-day Earth, and to end at a place where average nearest-neighbor
distances become larger than $10^4$~AU. After illustrating results for these
choices in sections 3.3 and 3.4, we return to the general case in      
\S 3.5. 

In \gc s, light from
other stars can provide a significant amount of energy.
 The ambient stellar flux is therefore of interest when considering
the opportunities available to advanced civilizations in \gc s. 
This is the total flux provided by the combination of all cluster stars.
We also want to know the 
flux provided by the brightest nearby star. The average
incident flux and the maximum received from a single star both
 are largest in regions where $D$ is small. That is, in regions where the
distance between nearest neighbors is smallest, planets may also be able to draw
energy from stars they do not orbit.   
This is also true for \ffp s, where energy drawn from nearby stars could
help to fuel any life they may harbor.  

\subsection{Method}

To conduct a quantitative search for the sweet spot,
we modeled both the cluster and its stellar population.
We used Plummer models for the cluster, because they are simply
characterized by a total cluster mass
$M$ and by a characteristic radius, $r_0$. They
allow us to derive analytic expressions for
the mass interior to each radius, and the average local speed, $v,$
as a function of radius.
To model the stellar population, we proceeded as follows.

We selected the initial
stellar population
from a Miller-Scalo initial-mass function (IMF) \citet{1979ApJS...41..513M}, 
considering all stars with masses above
$0.08\, M_\odot$.
We took the mass of the present-day turn off to be $0.84\, M_\odot$;
stars with slightly higher mass (up to $0.85\, M_\odot$), were considered
to be giants. Any star with an initially higher mass
 was considered to be a present-day stellar remnant:
we included $[0.6\, M_\odot$ white dwarfs, $1.4\, M_\odot$ neutron stars,
and $7\, M_\odot$ black holes] derived from stars with initial masses of
$[0.85\, M_\odot <M(0)< 8.5\, M_\odot, 8.5\, M_\odot < M(0) <35\, M_\odot,
M(0)>35\, M_\odot]$, respectively.
This produces a population in which $84.4\%$ of the stars are
main-sequence dwarf stars, $14.8\%$ of the stars are white dwarfs,
0.2\% are giants, and the remaining stars are primarily neutron stars.
The sizes of globular-cluster populations of
neutron stars and black holes is difficult to predict,
and only a small fraction of these compact objects can be discovered
through their actions as
X-ray binaries or recycled pulsars.

Compact objects make negligible contributions to the average flux,
and their presence doesn't alter the average distances between stars.
There are exceptions, when a compact object accretes matter from a close 
companion, emitting X-rays. The brightest X-ray binaries in Galactic
\gc s (low-mass X-ray binaries, LMXBs) typically have luminosities of $10^{36}-10^{37}$~erg~s$^{-1}$.
Their influence on any life associated with other stars is likely to 
be limited
because (1)~the numbers of bright LMXBs are small, with
$15$ known in the Milky Way's system of
\gc s \citep{2010AIPC.1314..135H}; 
(2)~many have low duty cycles; (3)~only \pl s relatively
near to even a bright LMXB receive more light from it than from their 
host star.
Furthermore, LMXBs tend to be
in or near the cluster core where survival of 
habitable-zone
\pl s would be challenging even in the absence of LMXBs,
especially for the higher-mass stars that
tend to be found there.
The successors to LMXBs are recycled millisecond pulsars which can also be
luminous. We will discuss them in \S 5.

We placed each star at a specific
randomly chosen point within $10\, r_0$ of the cluster center.
We kept track of the total mass of stars that should be
(according to the Plummer model)
within each shell of thickness $dr,$ and stopped adding stars to a
shell when it reached the appropriate mass. In this way the total mass
generated by each simulation matched the total mass we had selected for the Plummer model,
and the mass profile matched the analytically-derived cluster
profile.

We generated the luminosity of each star as follows.
For main-sequence stars with masses below $0.43\, M_\odot,$ we
set $l=0.23\, m^{2.3},$ where $l$ is the luminosity of the star and
$m$ is its mass. For main-sequence stars of higher mass, we used
$l= m^4$.  Giants 
(compact objects) were arbitrarily assigned $100\, L_\odot$
($0.001\, L_\odot$).

With each star assigned a position and luminosity, we computed the flux as
a function of $r$ by choosing a random point in each of $1000$ spherical
shells, centered on the cluster, with radii extending to $10\, r_0.$
At each randomly-selected point we computed the total flux provided by
all of the cluster stars, as well as the flux of the star that contributed
the most to the total flux reaching that point. We also computed the distance
from each randomly selected point to the nearest star.
The results are shown in Figures~1 and 2, where we have computed moving
averages for $D(r)$ and ${\cal F}(r)$

\subsection{Results}

We conducted a full set of calculations for two different cluster masses
($10^5\, M_\odot$ and $10^6\, M_\odot$). For each mass, we used three
separate values of $r_0: 0.1$~pc, $0.3$~pc, and $0.8$~pc. 
Figures 1 and 2  
each show results of calculations for a \gc \ with mass
$M=1\times 10^5\ M_\odot.$ They illustrate the trends we sought to explore.  
In Figure~1 (Figure~2) 
$r_0= 0.9$~pc ($r_0=0.1$~pc),  corresponding to low (high) concentrations. 
In the upper panel of each figure, the 
the logarithm to the base ten of the survival time is
plotted versus the logarithm to the base ten of $r/r_0.$
The upper (lower) curve corresponds to orbits in the \hz \ of a main-sequence
star of mass $0.1\, M_\odot$ ($0.8\, M_\odot$)\footnote{
For clusters with different values of $M,$ values of $\tau$ 
scale by a factor equal to the inverse  square-root of the ratio of the masses.}.
A horizontal line at $log(\tau)=9.65$ is plotted to
correspond to a survival time roughly equal to the present-day age 
of the Earth.

The globular cluster considered in Figure 1 has such a low concentration 
that \pl s in the \hz s of stars with masses below roughly 
$0.4\, M_\odot$  can survive throughout the cluster, even near the 
center. This is indicated by the red vertical line near $r=0$, and 
the red arrow pointing toward larger values of $r$. The blue vertical 
line and it's associated arrow indicate that survival near the 
center and throughout the cluster is also possible for planets in 
the habitable zones of $0.6\, M_\odot$ white dwarfs
\citep{2011ApJ...731L..31A}. For \pl s in the \hz s of dwarf 
stars with masses of $0.8\, M_\odot$, however, survival is possible 
only at somewhat larger values of $r$, as indicated by the orange line
with the rightward pointing arrow. 

In low-concentration clusters, the stellar density near 
the cluster's outer edge can be comparable to the stellar density 
in the vicinity of our Sun.  The globular cluster opportunity 
is therefore lost at large values of $r$. In the middle panel, we 
have drawn a horizontal line corresponding to interstellar 
nearest-neighbor distances of $10^4$~AU. 
If we rather arbitrarily posit that, for interstellar distances larger than 
this, the \gco \ is lost, then the ``sweet spot'' ends at the value of $r/r_0$
shown with an orange line and leftward pointing arrow in the middle
panel of Figure~2.    
This example illustrates that for, globular clusters with 
low central concentrations, the globular cluster sweet spot is a large
spherical shell.

Figure 2 shows the results for a globular cluster with a higher 
central concentration. In this case, survival of habitable-zone 
planets is possible only for larger values of $r$ than when
the stellar concentration is low. 
On the other hand, the stellar 
density remains high even as one approaches the cluster's edge. Thus,
the overall effect is that the sweet spot moves outward as the
stellar concentration increases.
Note that the edge of the sweet spot occurs at higher values of $r$ than 
those shown here.

Real globular clusters exhibit a phenomenon known as mass 
segregation, in which more massive stars tend to be over-represented 
in the cluster's central regions, while low-mass stars are 
over-represented in the outer regions. This effect could tend to 
place G dwarfs near the cluster center, where planets in their habitable 
zones are able to survive for only short times. Mass 
segregation may also place dwarf stars of the lowest masses near 
the outer edges of clusters. Thus, many low-mass stars
and \ffp s in 
low-concentration   clusters  may inhabit portions of the clusters 
where interstellar distances are large.  To explore this effect we 
conducted  simulations that modeled mass segregation.
Because the factors that determine the distribution
of masses within \gc s are complex and dynamical in nature, we have
employed a toy model, described below, which shows the general 
effect of mass segregation.   

At each value of $r$ we chose some of the stars from the Miller-Scalo
IMF, and some from a uniform distribution. To mimic mass
segregation, 
we favored the uniform distribution
near the center of the 
cluster, and the Miller-Scalo IMF toward the outer
edges. Specifically, when our simulated location was a distance $r$
from the cluster center, we generated a random number; if its value was
smaller than than $[1-r/(10\, r_0)]$, we chose from a uniform distribution,
 otherwise we chose from the 
Miller-Scalo distribution. For the more condensed cluster,
$85\%$ ($27\%$)  of stars of $0.1\, M_\odot$ ($0.8\, M_\odot$)
occupy the ``sweet spot''
for stars of that mass. 
Thus, the ``sweet spots'' not only exist, but are occupied.
For the less condensed cluster, the effects of mass
segregation were small   
on the low-$r$ side of the sweet spot, but the cut off at large $r$ 
 meant that only roughly $40\%$ of G dwarfs, and
$15\%$ of M and K dwarfs are in the ``sweet spot'' at any given time.   
It is therefore important to note that
 stars move throughout the cluster. Thus, 
because low-concentration gives habitable-zone orbits very long lifetimes,  
most stars 
 will spend a significant amount of time passing through the ``sweet spot'',
where stellar densities are high enough to decrease interstellar
travel times significantly.

\subsection{Stellar Habitable Zones, Globular Cluster Habitable Zones,
and the Sweet Spot}

Individual stars have habitable zones: regions in which it is neither too
hot nor too cold to allow liquid water to exist on the surface of an
Earth-like \pl . The locations of \hz\ boundaries are not fixed numbers. 
This is not only because life may exist under a wider range of conditions than
we know, but more specifically because several properties of stars, \pl s and
\pl ary orbits play roles in determining habitability.  
In the previous two sections, we have explored 
whether
the orbits of habitable-zone \pl s can survive in a dense \gc\ environment.
If the ambient stellar density is too high,
passing stars are likely to steal, destroy, or eject the \pl s that
had been in the \hz\ over time intervals too short for life to develop.
The radius at which the stellar density drops to the point that
a \pl \ in the \hz\ of a star of mass $m$ can survive, marks the
beginning of what can be called 
the {\sl \gc \ \hz\  (GC-HZ)} for stars of that mass. 
The GC-HZ extends from that point outward to the edge of the cluster.  
The characteristics of the GC-HZ are the following.  

\noindent The location of the inner edge of the GC-HZ (i.e., the inner edge of the
sweet spot) 

(1)~depends on the survival time considered.

(2)~has a strong dependence on stellar mass, because 
the stellar \hz\ is smaller for low-mass stars and interactions
are less likely to disrupt small orbits.

(3)~depends on the orbit of the star within the
cluster.
Stars within a \gc\ can travel from its central to its outer
regions. Whether a specific planetary orbit survives depends on how
much time the \ps\ spends at different distances, $r$, from the center
of the \gc .

The sweet spot incorporates a new concept, which is that the distances
between stars can play a role in the long-term survivability of an
advanced civilization. Large interstellar distances,
such as those common in the Solar neighborhood, imply long two-way
communication and interstellar travel times. It seems
likely that, if
interstellar distances are smaller by more than an order of magnitude,
the time needed to establish independent outposts would also be shorter.
The factor by which times need to be shorter is, at present, a matter of
conjecture. If disk civilizations live long enough on average 
to   establish
outposts, then the factor may simply be unity. 
Here we have assumed that a decrease in travel time (hence in $D$) by
a factor of $10$, provides any \ac s in \gc s with a stronger
opportunity to establish outposts. A limit on the value
of $D$, the  nearest-neighbor distance, determines the
location of the outer edge of the sweet spot.

The value of $D$, the nearest-neighbor distance,
 in \gc s versus its value in the Galactic disk
allows us to relate the travel times required in these  two  
environments. The nearest habitable \pl \ in both cases may well
be associated with another, more distant neighbor.
In fact, it may be the case, in both the Solar neighborhood and in \gc s, that
the density of \ps s in which members of a particular \ac \ could find
or make suitable habitats is significantly smaller than the overall
stellar density. As long as the relative density of suitable habitats
is either the same or larger in \gc s, then the travel times to
these suitable habitats would still be 
shorter than in the Solar neighborhood.

\subsection{Free-floating Planets}

Interactions involving
\ps s can have a range of outcomes.
Some \pl s are exchanged into other
\ps s or else ejected as a result of interactions. 
Wide-orbit planets
are more likely to be lost through interactions with passing stars.
Furthermore, when their release velocities are comparable to their
previous orbital velocities, the speeds are not generally large enough
to allow escape from the cluster.

Thus, \pl s around stars are constantly being stripped away
and joining the ranks of \ffp s, many of which remain bound to the cluster.
\Ffp s may be found in every part of the cluster, but many will be
ejected from regions near the center of the \gc . As they move away,
they will receive a
large but decreasing amount of energy from the star that had been their host. The bottom
panels of Figures 1 and 2 show that they will also continue to receive
significant flux from the other cluster stars. This flux will be
most significant in high-concentration clusters. Interestingly 
enough, in many cases, the dominant source of ambient light will be a single
star not previously related to the \ffp .   
 
If \ffp s are to house 
life, they must have outer layers that shield the life from fluxes 
of comets and asteroids, from high-energy particles, 
and from some portions of the electromagnetic spectrum. [See, e.g.,
 \citet{2011Icar..216..485B}.]
In our own 
Solar System, some moons of the outer planets are covered by ice 
that can serve as a shield for oceans \citep{2007Icar..189..424H}. 
Furthermore, since life requires energy, any life on free-floating planets must 
have sources of energy independent of irradiation by a single star.  
We are now coming to understand that there are myriad sources of 
energy on which moons of the outer Solar System can draw, ranging radioactivity
to tidal interactions. Some of these could also be available to \ffp s in \gc s.
Free-floating planets in globular clusters may be able to draw upon 
energy emitted by nearby stars.  It is for this reason that we have 
included the lower panels of Figures 1 and 2, which depict the 
average flux received as a function of distance from the cluster's 
center, as well as the flux received  from the brightest nearby star.

We have no information at present about \ffp s in \gc s, but
studies of both \sfr s \citep{2012ApJ...756...24S} and 
microlensing events \citep{2011Natur.473..349S} 
provide evidence for them in the disk.

Free-floating \pl s must be considered as   possible
habitats for life and for \ac s. \Ffp s also interact with
\ps s.   
The expression for the rate of interactions is the same as the 
rate for interactions with stars (\S 3.1). The distances 
of closest approach associated with dramatic effects however,  
tend to be smaller, yielding a smaller value for the rate of interactions 
per free-floating planet.  
To compute the  total numbers of interactions with \ffp s, we need to 
know their density. 
If their numbers are larger than the number of stars, the density
may also be larger. However, mass segregation 
is an important feature of globular clusters, and free-floating 
planets have masses much smaller than those of the cluster 
stars. We therefore expect that, even though the cluster's core may be
the point of origin of many free-floating planets,  
they may spend the large majority  of their time in the 
outer portions of the globular cluster. This means that, in spite 
of the large numbers of free-floating planets, their local spatial 
densities may tend to be low. 

That any life on \ffp s may need shielding by an outer layer 
at all times, could promote survival when asteroids strike, or when the
\ffp\ passes through a regions with a high flux of radiation (e.g., 
near an LMXB or close to a normal star).   
Thus, if there is life on \ffp s, 
it may be able to survive throughout the cluster. There would be no
inner boundary to the sweet spot for \ffp s. (That is, for \ffp s,
$R_{\rm low}^{\rm sweet}= 0.$) The outer edge of the sweet spot
 is determined by the same
considerations as for bound \pl s: the value of $D$ should be less than
some value, which we have taken to be $10^4$~AU.

There would be an interesting corollary should it be that 
\ffp s (1)~exist in \gc s, (2)~are more numerous than
bound \pl s, and (3)~are able to support life: 
the most meaningful nearest-neighbor distance    
would be the distance to the nearest \ffp . The
outer boundary of the sweet spot for both bound and \ffp s 
would move to larger values of $R_{\rm high}^{\rm sweet}$.
 
\begin{figure}[H]
\centering
\includegraphics[width=0.7\textwidth, height=160mm]{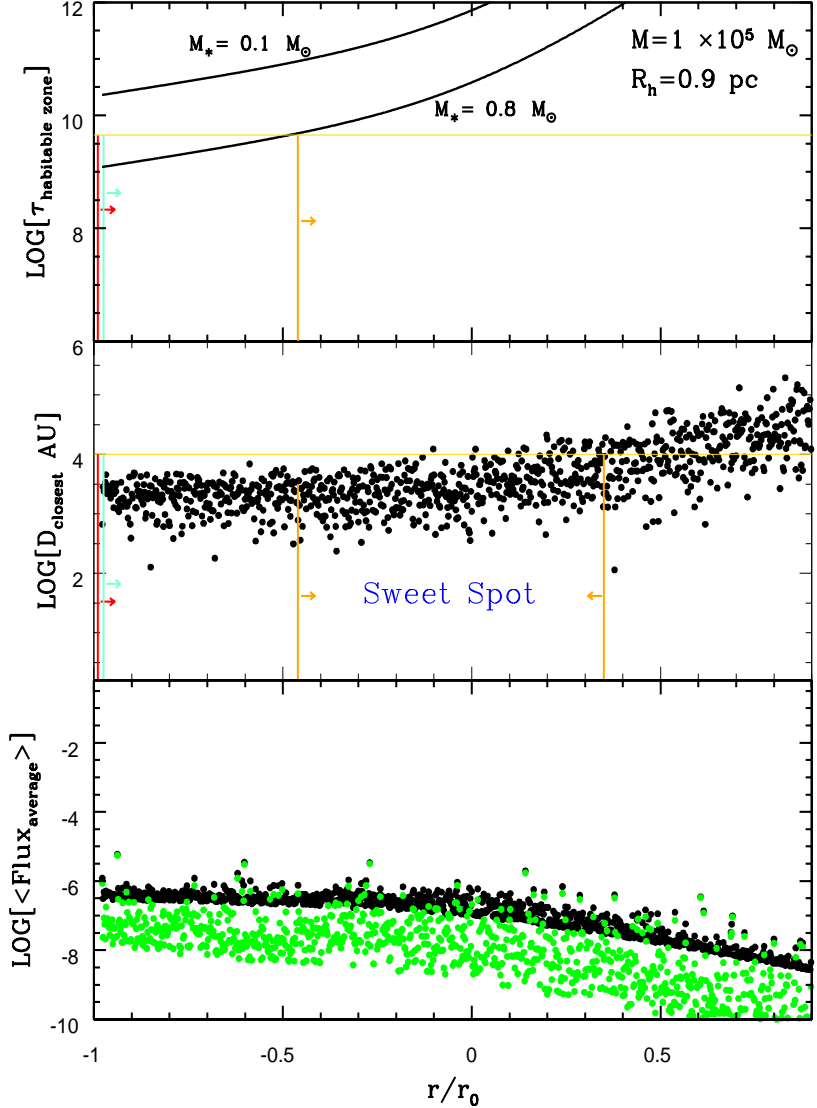}
\caption{\footnotesize 
{\bf A \gc \ with $M=10^5\, M_\odot$ and with 
relatively low concentration ($r_0 = 0.9$~pc).}
In the {\sl top panel} the logarithm (to the base 10) of the orbital
lifetime for a habitable-zone \pl \ is plotted against 
the logarithm (to the base 10) of $r/r_0$.
$\tau$ has different values for different stellar masses. 
Here, ($m_\ast=0.1\, M_\odot$, $m_\ast=0.8\, M_\odot$) in the
 (upper, lower) curve. In the {\sl middle panel} the logarithm of
the distance
$D_{closest}$ to the
nearest star
is shown.
In the {\sl bottom panel} the logarithm of the 
total flux received is plotted in
black. In green is the flux provided by the single star that provides the
most flux. 
The region marked ``sweet spot'' has values of $r/r_0$ high enough
and densities low enough that a planet in the habitable zone of
an $0.8\, M_\odot$ main-sequence star can survive; it also has
values of $r/r_0$ low enough and densities high enough that
nearest-neighbor distances are smaller than $10^4$~AU. 
The sweet spots for $0.1\, M_\odot$ stars (red) 
and also for white dwarfs
of mass $0.6\, M_\odot$ (aquamarine),  
end at the same place, but start at the lower values 
of $r/r_0$ indicated by the right-pointing arrows near the
left vertical axis. 
}
\end{figure}

\begin{figure}[H]
\centering
\includegraphics[width=0.7\textwidth, height=160mm]{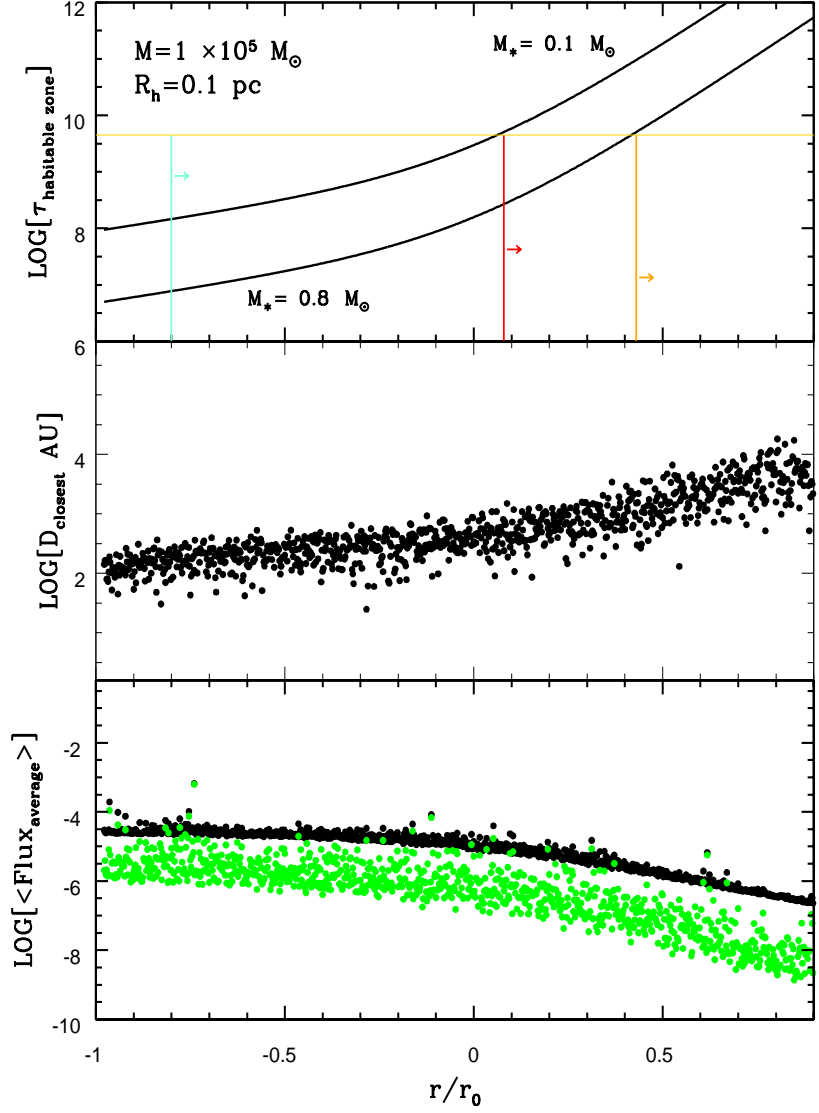}
\caption{\footnotesize
Same as in Figure 1, but with $r_0=0.1\, {\rm pc}$.
This is a more highly concentrated \gc . The sweet spots for each type of
star 
($0.8\, M_\odot$, orange; $0.1\, M_\odot$, red; $0.6\, M_\odot$, aquamarine)
each start at higher values of $r/r_0.$ But they extend to the highest value
of $r/r_0$ shown here. Of course at the very edge of a \gc , stellar densities
decline to the point that distances between stars are $>10^4$~AU, and we would
say that the sweet spots have ended.     
}
\end{figure}

\section{SETI and the \GC\ Opportunity}

Searches for extraterrestrial intelligence (SETI) started in the
1950s and 1960s \citep{1959Natur.184..844C, 1961PhT....14...40D,
1960Sci...131.1667D}, and references in
\citet{2001ARA&A..39..511T}, 
before planets beyond the Solar System had been discovered.
Today we  know of more than 2000 exoplanets (see, e.g., {\sl exoplanets.edu})
 but there are 
many open questions about planets, the formation of life,
the nature of intelligence, and the development and
lifetime of advanced civilizations.

Let $L_i$ represent the lifetime of an advanced civilization.
Our premise is that, once a \gc\ civilization is able to set up 
independent outposts, the probability becomes smaller
 that a catastrophic event will
eliminate all \ac s descended from it. 
We also postulate that 
smaller interstellar distances decrease the time $T_i$ it takes for a 
civilization to establish outposts.

These ideas incorporate several assumptions.   
First, interstellar travel must be possible in globular clusters. 
For example, the danger 
of impacts from small 
masses in interstellar space must not be too great.
Second, it must be possible to establish outposts. 
If, for example, 
life is plentiful, but incompatible in different planetary systems, 
it may be difficult to find hospitable environments. 
If both interstellar travel and the establishment of outposts can occur, 
 it is reasonable to consider that smaller 
interstellar distances could allow the first self-sustaining outpost 
to be established by a globular-cluster civilization at a time $T_i < L_i.$

\subsection{The Drake Equation}

The Drake equation, developed during the early years of SETI, 
identifies 
the factors that determine the
number of communicating civilizations in 
existence in the Galaxy at a typical time.
See, e.g., \citet{2008ASPC..395..213D}.
There are many possible definitions of the term ``communicating 
civilization". To set a scale, we will classify Earth as 
a planet with a communicating civilization, with a lifetime $L$, so far, 
of 100 years.

The
form most suitable for our purposes is the following,
where ${\cal N}_{\rm b}$ is the number of \cc s
on planets bound to stars.
\begin{align*} 
{\cal N}_{\rm b}
&  = N_\ast \times f_{\rm b}({\rm star|pl}) 
\times  n_{b}({\rm pl}) \times
f_{\rm b}({\rm pl|life}) \times  
f_{\rm b}({\rm life|int}) \times 
f_{\rm b}({\rm int|com}) \times \frac{L_{\rm b}}{\tau_{\rm H}}  \\
& = N_\ast \times {\cal F}_{\rm b} \times \frac{L_{\rm b}}{\tau_{\rm H}}  \numberthis
\end{align*} 
$N_\ast$ is the total number of stars in the disk.
$f_{\rm b}({\rm star|pl})$ is the fraction of stars with planets, and
$n_{\rm b}({\rm pl})$ is the average number of planets per star. The fraction
of planets on which life develops and the fraction of these on which
intelligent life develops, and the fraction of these on which \cc s
develop are,
respectively,
$f_{\rm b}({\rm pl|life})$, $f_{\rm b}({\rm life|int})$
and $f_{\rm b}({\rm int|com})$. These factors are combined
to form the overall factor ${\cal F}_{\rm b},$ the average number of
\cc s formed per star. 
The number of \cc s orbiting stars at any given time is proportional to
$L_{\rm b},$
the average lifetime of those communicating civilizations orbiting stars.
The ratio $L_{\rm b}/\tau_{{\rm H}}$ is the fraction of a Hubble time
over which there is a communicating civilization. 
 
\subsection{Comparing \GC s to the Galaxy}

\noindent{\bf Numbers of stars:}  
The first element of our comparison is the ratio 
$R_\ast = N_\ast^{\rm gc}/N_\ast$, where the
numerator is the number of stars in a \gc .
This varies among \gc s from under $10^5$ to more than $10^6$.
\begin{equation} 
R_\ast = 5 \times 10^{-6} 
\Bigg(\frac{N_\ast^{\rm gc}}{5\times 10^5}\Bigg) 
\Bigg(\frac{1 \times 10^{11}}{N_{\ast}}\Bigg) 
\end{equation} 

If all other factors were equal, a population of a few times $10^3$  
communicating civilizations in the disk would correspond to $\sim 1$ in
the Galaxy's population of $\sim 150$ globular clusters. 
A disk population roughly 100 times as large
would correspond to $\sim 1$ \cc \ within each of many \gc s.   
\smallskip

\noindent{\bf Value of ${\cal F}_{\rm b}$:}  
The second element of our comparison is the factor ${\cal F}_{\rm b}$, the number
of \cc s formed per star. The fact that \gc\ stars are long-lived 
means that a large fraction of them provide environments stable enough
for life to form and evolve on their planets. 
In fact,  old planetary systems may have had several opportunities to produce 
civilizations during the past 12~Gyr. While 
the same is true for old stars in the disk, 
only a smaller fraction of them are as old as globular cluster 
stars. 

In addition, evolution on \gc \ \pl s is less likely to be 
subject to interruptions. For example,  
astronomical events and excess exposure to radiation and winds can
essentially ``reset'' the clock for evolving life and civilizations.
These ``resets'' can delay evolution toward advanced civilizations, or
destroy them.
Because \gc s have little gas and dust, they
  do not form stars or produce
core-collapse supernovae or long gamma-ray bursts.

Of course, stellar passages have the effect of interrupting
developments on those planets that are either ejected because of an 
interaction, 
 or else
come to orbit another star after an interaction. We have shown, however, that
large numbers of \pl s in the \hz s of  low-mass stars should be
stable throughout  significant portions of most \gc s. From the perspective of
developing and evolving life in a manner that may parallel what 
happened on Earth, these are the most important systems, and this is why
${\cal F}_{\rm b}$ may have values in \gc s similar to those in the disk.

It is also important to consider, however, 
that orbits of  many planets are disrupted through interactions.
Ejection is especially   
likely for \pl s in wider orbits, where liquid water
could not have been be sustained. 
Ejections transform these \pl s into free floaters.
Life that  had existed on a planet losing its star, could
expire and/or develop
differently afterward. 
 \footnote{There are other questions 
to be considered. For example,
it would be difficult at present to assess the relative 
frequencies of asteroid
strikes in \gc s versus the disk. Stars in \gc s cannot 
have extended asteroidal disks or clouds. This would tend to
 decrease the frequency of impacts.
 On the other hand, the ambient density of planetoids may be higher
in \gc s. 
Nevertheless, because asteroids and comets have very small masses,
 they would tend to migrate toward the outer
edges of a \gc , helping to moderate the average  
density throughout much of the cluster. A second question is
the rate of Type Ia supernovae. Observations have a established 
 that they do not occur more frequently in globular clusters than 
in the field \citep{2013IAUS..281...21V, 2013ApJ...762....1W}; 
these numbers will continue to be refined.} Finally, changes in orbit
can be induced by stellar passages, even when a \pl ary 
system's architecture is
not reconfigured. These effects, though more modest,
could nevertheless influence life 
in a \ps \
(in either a positive or negative way), and must  be considered
in more detailed work, similar to the way  issues such as orbital eccentricity 
are now considered in computations involving the habitable zone.  

\smallskip 

\noindent The enhanced stability of the \gc\ environment is
part of what we can call the {\sl globular cluster opportunity}. 

\smallskip 

\noindent{\bf Lifetimes of \cc s:} 
The final factor is the lifetime of \cc s. 
Let $Y_i(p)$ be the number of \ps s in which one or more \cc s develop.
We can classify a \cc \ according to whether it 
remains within a single  \ps, or whether it develops outposts 
outside of it. 
\begin{equation}  
\frac{L}{\tau_H} = \frac{1}{\tau_H} \times  
\sum\limits_{i=1}^{Y(p)} 
\, \Bigg[\sum\limits_{j=1}^{C_i^1} L_i 
+ 
\sum\limits_{j=1}^{C_i^{>1}} \Big(\tau_H-t_i(0)\Big)\times \eta_i\Bigg]   
\end{equation}  

The outer summation in Eq.~4 is over planetary systems, the 
inner summation 
is over the sequence of civilizations in a given \ps . 
The first term 
represents 
communicating civilizations that do not establish outposts. 
The second term 
represents 
communicating civilizations that do establish outposts. 
Once a set of independent, self-sustaining outposts has been
established, the cluster may always host descendants of the 
original ``seed'' civilization. 
Thus, the value of $L$
is simply the difference between the present time and 
the start of the seed civilization. We include the
factor $\eta < 1$ to recognize that effects we cannot anticipate
may lead to the end of these civilizations 
in spite of the apparent opportunity to continue into the indefinite future.

\smallskip

\noindent The second part of the {\sl globular cluster opportunity} is 
that relatively small interstellar distances may allow
self-sustaining outposts to be developed 
over relatively short time scales. This would give \gc s 
the potential to host \cc s over a continuous
very-long-lasting epoch.

\smallskip

\subsection{Conditions for \GC \ Civilizations}

Our discussion of the Drake equation has focused on the numbers of communicating civilization expected at present. It is useful to also consider
the total number $n$ of civilizations that are ever formed
within a stellar population (either a galaxy or a \gc ).
\begin{equation}  
n = N_\ast \times {\cal F} 
\end{equation}  
Let's suppose that, over the course of a Hubble time, a certain
minimum number, $n_{\rm min}$, of \cc s
 must arise within a specific \gc \ in order to  
ensure that one of them will be
able to establish 
self-sustaining outposts. In order for this minimum value to be achieved, 
the value of ${\cal F}^{\rm gc}$ must be greater than 
${\cal F}_{min}^{\rm gc} =n_{min}^{\rm gc}/N_\ast^{\rm gc}.$

\begin{equation}     
{\cal F}_{min}^{\rm gc} = 
10^{-5} \times \Big(\frac{n_{min}^{\rm gc}}{10}\Big)\, 
\Big(\frac{10^6}{N^{\rm gc}_\ast}\Big)
\end{equation}     
This equality translates a minimum value of $n$ into a minimum
value of ${\cal F}$, which can then be related, through 
Equation~2 into a condition on the factors whose product is $\cal F$.

For example, one way to achieve a value ${\cal F}_{min}^{\rm gc}=10^{-5}$ 
is if only $10\%$ of
 cluster stars have \pl s that can 
support life, only $1\%$ of \pl s with life 
support intelligent life, and $1\%$ of planets with intelligent life  
produce \cc s. 
These relatively low probabilities could be enough to ensure that
every \gc\ hosts a long-lived \cc , even if only 
one in ten \gc\ \cc s succeeds in establishing outposts.

To place these values in context, we consider  
the galactic disk. Should ${\cal F}^{gal}$ 
be as small as $10^{-5}$, then $10^{11}$ disk stars would
produce $10^6$ \cc s in a Hubble time. If these  
each last for a time $10^{3+k}$~years, with $k$ ranging from 1 to 7,
there would be $10^{k-1}$ \cc s at any one time. In the small-$k$
limit, we could  be the only \cc\ active in the Galaxy today. In the 
large-$k$ limit, the nearest \cc\ would be on the order of $100$~pc away.
\footnote{This distance could be reduced, however, if galactic \cc s 
produce self-sustaining outposts, and/or if \gc\ \cc s spread to the
disk.}   

This example illustrates  
that 
many of the 
Milky Way's \gc s could presently
host advanced \cc s that have spread throughout the 
cluster, whether the disk of the Galaxy contains no other \cc s or
whether it is rich in \cc s. 
Furthermore, if \gc s do host \ac s, they will tend to be
old civilizations.
 
\subsection{Free-floating \pl s}

The Drake Equation can be applied to \ffp s.
Let ${\cal N}_{\rm f}$ be the number of \cc s on \ffp s.  
\begin{align*} 
{\cal N}_{\rm f}
&  = N_{\rm f} \times 
f_{\rm f}({\rm pl|life}) \times  
f_{\rm f}({\rm life|int}) \times 
f_{\rm f}({\rm int|com}) \times \frac{L_{\rm f}}{\tau_{\rm H}}  \\
& = N_{\rm f} \times {\cal F}_{\rm f} \times \frac{L_{\rm f}}{\tau_{\rm H}}  \numberthis
\end{align*} 
$N_{\rm f}$ is the total number of \ffp s in the disk.
The fraction of \ffp s on which life develops and the fraction of these on which
intelligent life develops, and the fraction of these on which \cc s
develop are,
respectively,
$f_{\rm f}({\rm pl|life})$, $f_{\rm f}({\rm life|int})$
and $f_{\rm f}({\rm int|com})$. These factors are combined
to form the overall factor ${\cal F}_{\rm f},$ the number of
\cc s formed per \ffp .
$L_{\rm f},$
is the average lifetime of those communicating civilizations on \ffp s.

We don't know the number of \ffp s. Based on microlensing surveys,
the disk population of \ffp s appears to be larger than the disk population
of main-sequence stars \citep{2012MNRAS.423.1856S, 2011Natur.473..349S}.  
Setting $N_{\rm f}$ to
$\phi \times N_\ast,$ the value of ${\cal F}_{\rm f}$ 
needed to produce a certain
number of \cc s is proportional to $1/\phi$.

Thus, large values of $\phi$ mean that  
the value of ${\cal F}_{\rm f}$ can be even smaller than
${\cal F}_{\rm b}$, to produce a number of \cc s on \ffp s
comparable to the number of \cc s on \pl s bound to stars. 
Put another way, the chance of a \cc \ developing on an \ffp \ can be
very small, making  
life extremely uncommon among \ffp s; yet 
there may be more \cc s developing on \ffp s
than on \pl s in the \hz s of stars.
\Ffp s in \gc s have an advantage, in that the stars in their vicinity may
provide significant energy. This is especially so if the civilization is
advanced enough to build and transport large stellar-light collectors.

\section{Implications} 

\subsection{Overview} 
Although only a single \pl\ has so far been discovered in a \gc , 
several lines of reasoning suggest that globular-cluster 
planets may be common.
If there is a  similarity to \ps s in the disk, then low-mass 
cluster stars may host \pl s in their \hz s. We have shown that
there are large regions of \gc s, ``sweet spots'',
 in which (1)~habitable-zone \pl ary orbits 
have long lifetimes,
while (2)~the distances between neighboring stars are small 
enough to significantly decrease
interstellar travel times from what they are in the Galactic disk.  

The existence of a ``sweet spot'', possibly combined with long-term 
stability afforded by the lack of massive stars in \gc s, is what
we have referred to as the \gco .  
If life and \ac s develop on the habitable-zone planets, then it is
reasonable to consider the possibility that the
lifetime of some \gc\ civilizations may exceed the time needed
to establish independent outposts. Should this be the case,
then \gc s may host \cc s that are old and are spread
throughout the cluster.   

\subsection{Prioritized list of \gc s}

We aim to identify those \gc s most likely to have 
large sweet spots. 
If the primary criterion we needed to impose were the existence of
\pl s in the \hz s of cluster stars, this would favor low stellar densities.
We want, however, to also have relatively small distances between
neighboring stars, which favors high densities. 

We consider an analogy with LMXBs, and their 
progeny, recycled millisecond pulsars. 
The formation of LMXBs has been explained in terms of interactions made 
possible by a high-density environment
\citep{1975ApJ...199L.143C}. As a result of these interactions,
a neutron star comes to have a binary companion that will donate mass to
it. 
The interaction which led to the formation of this binary  
is likely to have involved one or more binaries
 in the initial state. Furthermore, 
before post-interaction mass transfer can start, 
the newly formed neutron star binary must generally survive 
for a significant length of time before the donor comes to fill
its Roche lobe.  
 These circumstances suggest that the cluster must include, not far from the core,   
regions of modest density.  In fact, the orbit of the planet in M4 has a large
semimajor axis ($\sim 23$~AU), which would not survive in the cluster core.

This balance of higher and lower density is similar to the qualities
we seek in a \gc\ that has a significant sweet spot. 
These points are demonstrated empirically in Figure~3, each of whose
points corresponds to a \gc \ whose parameters have been taken from the 
Harris (2010) catalog. Along the horizontal axis is  
the log of the central luminosity density, 
$\rho.$ 
Along the vertical axis is the 
logarithm to the base 10 of the {\sl half-mass concentration factor}, 
which we have defined to be $h=log_{10}(R_h/R_t)$.
We used this factor because the value of $R_h =1.3\, r_0$ 
(from the Plummer model) is directly tied to the overall fall-off of the 
cluster density with distance from the center.

In Figure~3, points with a yellow triangle superposed correspond to \gc s that
host \mspsr s. Those surrounded by green rings contain 
at least $3$ \mspsr s and those
surrounded by red rings have 
$10$ or more \mspsr s. The figure displays two trends.
First, there are no discovered \mspsr s in \gc s with
$log_{10}(\rho) < 2.8$.
Second, neither very high nor very low concentration factors are
associated with multiple \mspsr s. 
While the trends in Figure~3 almost certainly reflect 
observational selection effects convolved 
with physical principles (for example, each cluster may host more \mspsr s
than observed), they are consistent
with the results of \S 3 for habitable-zone \pl s illustrated in Figures~1
and 2.
In Table~1 we therefore use the numbers of \mspsr s as a proxy to help prioritize  
searches for planets and for intelligent life in \gc s, keeping in mind that
other factors may eventually be understood to be more important. 
The columns are the: cluster name; distance from us; metalicity;
concentration factor computed from the the ratio of the core to tidal radius;
core radius; half-mass radius; concentration factor, $h$, as defined above;
central luminosity density; and the number of known recycled pulsars.

\subsection{Navigation} 

Millisecond pulsars 
provide strong, stable, periodic signals, 
which in 19 of the \gc s listed in Table~1, 
emanate from different directions. 
Timing precision
allows the determination of their
positions with great accuracy and this precision has in turn led to suggestions
of using pulsar timing to navigate spacecraft [\citet{downs74},
see \citet{2013AdSpR..52.1602D} for a recent review].
That is, measured timing residuals of multiple
pulsars can be used to determine the spacecraft position
with a precision that depends on the accuracy of the measured
times of arrival of pulses (TOAs) and the stability of the
pulsars. Because pulsar observations using radio
telescopes require large collecting area of the telescopes
that are impractical for spacecrafts, X-ray observations
of pulsars using much smaller X-ray telescopes have
been proposed for spacecraft navigation \citep{1981TDAPR..63...22C}
(see also \citet{2006JGCD...29...49S} and US Patent 7197381B2).
With an ensemble of four millisecond
pulsars and realistic timing accuracy \citet{2013AdSpR..52.1602D} show
that the position of a spacecraft can be determined to
an accuracy of 20 km on a trajectory from Earth to Mars.
Since in a globular cluster environment a set of
pulsars will be within a typical distance $< 10$ pc, 
the pulsars appear far brighter than when viewed 
from the Solar System, thousands of 
parsecs away.   
The potential use of small radio antennae 
could allow {\it radio} pulsar
timing with the precision needed for navigation.

\subsection{Search for \pl s} 

The crowding of dim \gc\ stars,
at distances larger than a kpc (Table 1), makes it challenging
to discover globular-cluster \ps s. The progress made
during the past several years in discovering \pl s in open clusters is, however,
a positive development. Transit studies of the outer regions
of \gc s would allow us to focus on \pl s in the habitable zone while
taking advantage of mass segregation. 
The most numerous stars would be very low-mass M dwarfs,
and their small sizes would optimize the chances of
discovering the small planets that are expected.

High resolution studies like those conducted by 
\citet{2000ApJ...545L..47G} with HST, could be effective in regions 
of higher density. In the core, however, 
mass
segregation could
 mean that the most common main sequence stars are those of relatively 
high mass. Orbital periods of \pl s in the \hz \ could be
dozens or hundreds of days.
It would therefore be more productive to study dense regions located
outside the core. 
 Even so, the baselines would need to be 
long enough for the discovery of planets in orbits that have 
periods up to a few tens of days. 

Another important step would be 
to discover free floating planets. Discoveries of free floating planets 
in the field have been reported by microlensing teams
\citep{2011Natur.473..349S}.
 Microlensing is ideally suited for these discoveries, 
because gravitational lensing is sensitive to mass; 
light from the planet is not required.  As it happens, several 
globular clusters lie in fields studied by optical monitoring team. 
Excess events along the directions to these clusters 
have been reported \citep{2014AAS...22430001D, 2015mgm..conf.2075J}. The rate of lensing events due to globular 
cluster stars is expected to be small \citep{1994AcA....44..235P}. 
But
with the improved monitoring now being conducted, 
enough \gc\ lensing events will have been discovered 
that it should be possible 
to discover or place limits on free-floating planets in \gc s.
Furthermore, knowing the distance and proper motion of 
the cluster would allow the mass of each planet so discovered to be measured.

\subsection{SETI}
In 1974 a radio message was beamed from Aricebo to the \gc \ M13 
({\sl http://www.seti.org/seti-institute/project/details/arecibo-message}).
If that message is received and answered promptly, it will take almost 
42,000 years for us to receive a response. Although other \gc s are closer,
almost all are more than a kpc away, making short-term two-way communication
problematic. 

If, therefore, we are to find evidence of extraterrestrial intelligence 
in globular clusters, it will be through signals that originated 
in the clusters long ago. These signals may represent attempts at 
communication with \ac s in the 
Galactic disk. Or they may be signals generated incidentally as a \gc\ 
society
carries out its normal functions. With more than 50 years of work
on this topic, many
ideas have been developed \citep{2001ARA&A..39..511T}.

If \cc s are common in the  
Galaxy, \gc s may be good targets for SETI, simply because they are 
dense, well-defined stellar systems. 
In \S 4 we showed that, even if \cc s are rare in the disk of the Milky Way,
they could occupy multiple Galactic \gc s, 
and could be very advanced. Although discussions of
long-lived \ac s are necessarily speculative,
it may be easier to detect signals from
an advanced civilization. If the signals involve energetic phenomena, such as 
X-ray emission from LMXBs, they could be detectable even if they emanate from
\gc s outside the Milky Way. There are many thousands of \gc s within
$10$~Mpc. Radio emission from \gc s within the Milky Way is regularly studied,
and X-ray emission is studied from Milky Way \gc s and, 
at least on occasion, from thousands of
\gc s in galaxies as far from us as the Virgo cluster. These data are
studied with the goals of learning more about accreting
compact objects.  It may be worthwhile to enhance the analysis
for subtle additional signatures that could be signs of intelligent
life.

\begin{figure}
\includegraphics[scale=1.0]{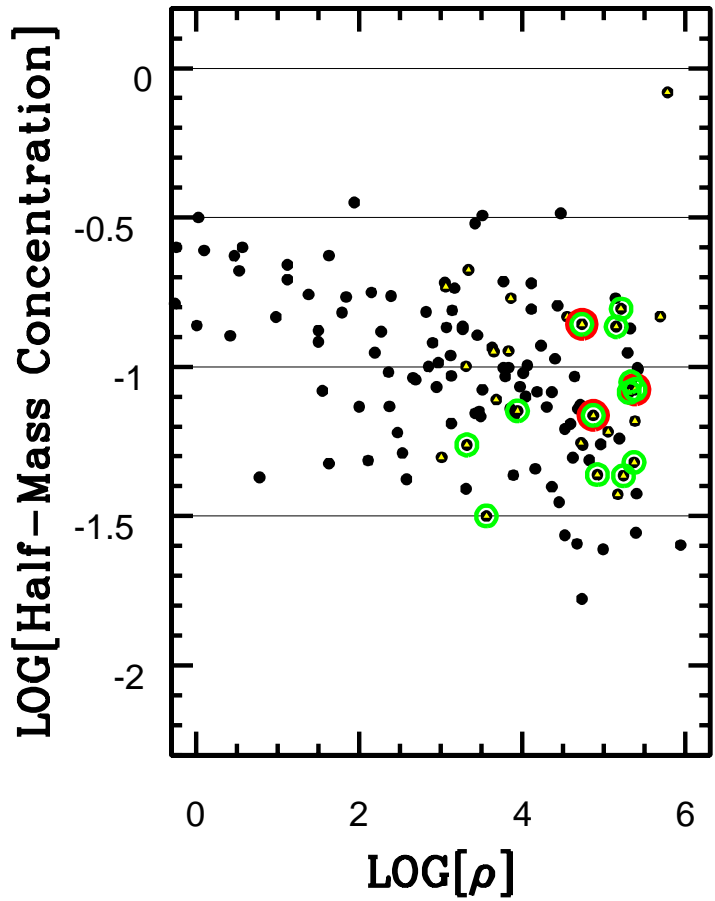}
\vspace{-5.5 true in}
\caption{
{\bf Logarithm (to the base ten) of the half-mass concentration factor, $h$, 
versus logarithm (to the base ten) of $\rho,$ which here is taken to be
the central luminosity density (Harris 2010).} 
Points with a yellow triangle superposed correspond to \gc s that
host \mspsr s. Those surrounded by green rings contain
at least $3$ \mspsr s and those
surrounded by red rings have
$10$ or more \mspsr s.} 
\end{figure}
\begin{table}
\centering
\caption{Globular Clusters Ordered by Numbers of Recycled Pulsars}
\label{my-label}
\begin{tabular}{lllllllll}
 {\bf Cluster} &  {\bf D (kpc)} & {\bf [Fe/H]} & {\bf c$_{\rm core}$} & {\bf R$_{\rm c}$} & {\bf R$_{1/2}$} & {\bf c$_{1/2}$} & {$\bf \rho$}                    & {\bf N$_{\rm msp}$} \\
Terzan5 &    8.0 &   -0.28 &    1.74 &    0.18 &    0.83 &    1.08 &    5.38 &    35  \\
NGC104  &    4.3 &   -0.76 &    2.04 &    0.37 &    2.79 &    1.16 &    4.87 &    23  \\
NGC6626 &    5.7 &   -1.45 &    1.67 &    0.24 &    1.56 &    0.86 &    4.73 &    12  \\
NGC7078 &   10.2 &   -2.22 &    2.50 &    0.07 &    1.06 &    1.32 &    5.37 &     8  \\
NGC6624 &    7.9 &   -0.42 &    2.50 &    0.06 &    0.82 &    1.36 &    5.24 &     6  \\
NGC6440 &    8.0 &   -0.34 &    1.70 &    0.13 &    0.58 &    1.05 &    5.33 &     6  \\
NGC6266 &    6.7 &   -1.29 &    1.70 &    0.18 &    1.23 &    0.87 &    5.15 &     6  \\
NGC6752 &    3.9 &   -1.55 &    2.50 &    0.17 &    2.34 &    1.36 &    4.92 &     5  \\
NGC6205 &    7.0 &   -1.54 &    1.49 &    0.88 &    1.49 &    1.26 &    3.32 &     5  \\
NGC5904 &    7.3 &   -1.29 &    1.87 &    0.40 &    2.11 &    1.15 &    3.94 &     5  \\
NGC6517 &   10.5 &   -1.37 &    1.82 &    0.06 &    0.62 &    0.81 &    5.21 &     4  \\
NGC6441 &    9.7 &   -0.53 &    1.85 &    0.11 &    0.64 &    1.09 &    5.31 &     4  \\
NGC5272 &   10.0 &   -1.57 &    1.85 &    0.50 &    1.12 &    1.50 &    3.56 &     4  \\
NGC6522 &    7.0 &   -1.52 &    2.50 &    0.05 &    1.04 &    1.18 &    5.38 &     3  \\
NGC7099 &    7.9 &   -2.12 &    2.50 &    0.06 &    1.15 &    1.22 &    5.05 &     2  \\
NGC6760 &    7.3 &   -0.52 &    1.59 &    0.33 &    2.18 &    0.77 &    3.86 &     2  \\
NGC6749 &    7.7 &   -1.60 &    0.83 &    0.77 &    1.10 &    0.68 &    3.34 &     2  \\
NGC6656 &    3.2 &   -1.64 &    1.31 &    1.42 &    3.26 &    0.95 &    3.65 &     2  \\
NGC6544 &    2.5 &   -1.56 &    1.63 &    0.05 &    1.77 &    0.08 &    5.78 &     2  \\
NGC6838 &    3.8 &   -0.73 &    1.15 &    0.63 &    1.65 &    0.73 &    3.06 &     1  \\
NGC6652 &    9.4 &   -0.96 &    1.80 &    0.07 &    0.65 &    0.83 &    4.55 &     1  \\
NGC6539 &    7.9 &   -0.66 &    1.60 &    0.54 &    1.67 &    1.11 &    3.68 &     1  \\
NGC6397 &    2.2 &   -1.95 &    2.50 &    0.05 &    2.33 &    0.83 &    5.69 &     1  \\
NGC6342 &    9.1 &   -0.65 &    2.50 &    0.05 &    0.88 &    1.25 &    4.72 &     1  \\
NGC6121 &    2.2 &   -1.20 &    1.59 &    0.83 &    3.65 &    0.95 &    3.83 &     1  \\
NGC5986 &   10.3 &   -1.67 &    1.22 &    0.63 &    1.05 &    1.00 &    3.31 &     1  \\
NGC5024 &   18.4 &   -2.07 &    1.78 &    0.37 &    1.11 &    1.30 &    3.01 &     1  \\
NGC1851 &   12.2 &   -1.26 &    2.24 &    0.08 &    0.52 &    1.43 &    5.17 &     1  \\
\end{tabular}
\end{table}

\begin{acknowledgements}
This research has made use of NASA's Astrophysics Data System.
RD would like to thank Kevin Hand for discussions.   
\end{acknowledgements}

\end{document}